\documentclass[
  aps,
  prl,
  reprint,
  superscriptaddress,
  amsmath,
  amssymb,
  nofootinbib,
  longbibliography,
]{revtex4-2}

\usepackage{amssymb}
\usepackage{lipsum}
\usepackage{graphicx}
\usepackage{multirow}
\usepackage{amsmath,amssymb,amsfonts}
\usepackage{amsthm}
\usepackage{mathrsfs}
\usepackage[title]{appendix}
\usepackage{textcomp}
\usepackage{booktabs}
\usepackage{algorithm}
\usepackage{algorithmicx}
\usepackage{algpseudocode}
\usepackage{listings}
\usepackage{xspace}
\usepackage{rotating}
\usepackage{siunitx}
\usepackage{orcidlink}

\DeclareSIUnit{\mrad}{mrad}

\newcommand*{\mrec}{\ensuremath{\rm m_{\text{recoil}}}\xspace}
\newcommand*{\mh}{\ensuremath{\rm m_{\text{H}}}\xspace}
\newcommand*{\mZ}{\ensuremath{\rm m_{\text{Z}}}\xspace}
\newcommand*{\mll}{\ensuremath{\rm m_{\ell\ell}}\xspace}

\newcommand*{\pll}{\ensuremath{p_{\ell\ell}}\xspace}
\newcommand*{\costhetamiss}{\ensuremath{\cos(\theta_{\text{miss}})}\xspace}

\newcommand*{\GeV}{\xspace\ensuremath{\text{Ge\kern-0.1em V}}\xspace}
\newcommand*{\MeV}{\xspace\ensuremath{\text{Me\kern-0.1em V}}\xspace}
\newcommand*{\ZH}{\ensuremath{\text{ZH}}\xspace}
\newcommand*{\eeZH}{\ensuremath{\mathrm{e^{+}e^{-}} \rightarrow \text{ZH}}\xspace}
\newcommand*{\iso}{\ensuremath{I_{\text{rel}}}\xspace}

\newcommand*{\Z}{\ensuremath{\text{Z}}\xspace} 
\newcommand*{\WW}{\ensuremath{\text{WW}}\xspace} 
\newcommand*{\ZZ}{\ensuremath{\text{ZZ}}\xspace}
\newcommand*{\Zg}{\ensuremath{\text{Z/}\gamma}\xspace}
\newcommand*{\invab}{\ensuremath{ab^{-1}}\xspace}

\newcommand*{\WHIZARDPs}{{\tt WHIZARD3+PYTHIA6}\xspace}

\newcommand*{\PYTHIAe}{{\tt PYTHIA8}\xspace}
\newcommand*{\DELPHES}{{\tt DELPHES}\xspace}
\newcommand*{\KEYfHEP}{{\tt Key4HEP}\xspace}

\newcommand*{\ZeeH}{\ensuremath{\text{Z}(\mathrm{e^{+}e^{-}})\text{H}}\xspace}
\newcommand*{\ZmumuH}{\ensuremath{\text{Z}(\mathrm{\mu^{+}\mu^{-}})\text{H}}\xspace}

\newcommand*{\sqrts}{\ensuremath{\sqrt{s}}\xspace}
\newcommand*{\sqrtsZH}{\ensuremath{\sqrts = 240~\GeV}\xspace}
\newcommand*{\sqrtsTop}{\ensuremath{\sqrts = 365~\GeV}\xspace}
\newcommand*{\sqrtsZHTop}{\ensuremath{\sqrts = 240\text{ and } 365~\GeV}\xspace}

\begin{document}

\title{Toward a Measurement of the Higgs Boson Mass\\ with Natural-Width Precision at FCC-ee}

\author{J.~Eysermans}
\affiliation{
Particle Physics Collaboration,
Massachusetts Institute of Technology,
77 Massachusetts Avenue,
Cambridge, Massachusetts 02139, USA
}

\author{A.~Li}
\affiliation{
Physics Department,
Brookhaven National Laboratory,
Upton, New York 11973, USA
}
\affiliation{
Laboratoire AstroParticule et Cosmologie,
CNRS/IN2P3,
10 rue Alice Domon et Léonie Duquet,
75013 Paris, France
}

\author{G.~Bernardi}
\affiliation{
Laboratoire AstroParticule et Cosmologie,
CNRS/IN2P3,
10 rue Alice Domon et Léonie Duquet,
75013 Paris, France
}

\author{C.~Paus}
\affiliation{
Particle Physics Collaboration,
Massachusetts Institute of Technology,
77 Massachusetts Avenue,
Cambridge, Massachusetts 02139, USA
}

\author{E.~Perez}
\affiliation{
CERN,
Esplanade des Particules 1,
1211 Geneva 23, Switzerland
}

\author{M.~Selvaggi}
\affiliation{
CERN,
Esplanade des Particules 1,
1211 Geneva 23, Switzerland
}

\begin{abstract}
Higgs boson mass measurements with sub-$10~\MeV$ precision enable
sub-percent determinations of Higgs boson couplings and prevent the Higgs
boson mass from becoming a limiting input to electroweak fits. Probing the
electron Yukawa coupling through resonant Higgs boson production requires a
precision comparable to the Higgs boson natural width of approximately
4~MeV. Using the leptonic $\ZH$ recoil channels, we show that FCC-ee can
reach a Higgs boson mass precision of 4~MeV, including statistical and
systematic uncertainties, thereby enabling this unique measurement. We
identify the detector and accelerator performance required to reach this
precision.
\end{abstract}

\maketitle

After the discovery of the Higgs boson by the ATLAS and CMS Collaborations in 2012~\cite{HIGG-2012-27, CMS-HIG-12-028}, with a measured mass of approximately 125\GeV, studies at the LHC have increasingly focused on precision measurements of its properties. Proposed future electron-positron colliders such as the Future Circular Collider (FCC-ee) will significantly extend this program, enabling determinations of Higgs boson properties with an order-of-magnitude improvement in precision~\cite{FCC:2025lpp}. In particular, the clean experimental environment enables measurements of Higgs boson couplings to hadrons with unprecedented precision ~\cite{DelVecchio:2025gzw}, while the precisely known initial state in electron-positron collisions allows for a model-independent determination of the total \ZH cross section at the few-per-mille level~\cite{li2025modelindependentzhproductioncross} and, consequently, of the absolute Higgs boson couplings and total width~\cite{morange_2023_ysabc-wm427}.

A precise determination of the Higgs boson mass, \mh, a free parameter of the Standard Model, is one of the central objectives of the FCC-ee program~\cite{Azzurri:2021nmy}. A precision of about 10\MeV or better is motivated by the dependence of the $\rm H\to WW^\ast$ and $\rm H\to ZZ^\ast$ partial widths on \mh through the available phase space, ensuring that the associated parametric uncertainties remain subdominant to the projected precision of Higgs boson coupling measurements. Beyond improving the determination of a fundamental parameter, a few-MeV measurement of \mh provides a decay-independent mass reference, renders the uncertainty in \mh negligible in global Higgs fits, establishes an absolute benchmark for exclusive Higgs reconstruction channels, and sets quantitative requirements on detector and accelerator performance, including tracking, electron reconstruction, beam-energy calibration, and beam-energy-spread control. Furthermore, FCC-ee uniquely allows for the study of direct resonant Higgs production in electron-positron collisions at a center-of-mass energy equal to $\rm \sqrt{s}=\mh$, providing sensitivity to the electron Yukawa coupling~\cite{dEnterria:2021xij,Fatehi:2026gnt}. Such a dedicated run requires prior knowledge of \mh with a precision comparable to, or better than, the Higgs boson natural width of about 4\MeV, thereby enabling a direct probe of the electron-Higgs interaction.

The current uncertainty in the Higgs boson mass is about 100\MeV, as measured by the ATLAS and CMS Collaborations~\cite{HIGG-2022-20, CMS-HIG-21-019}, using the $\rm \mathrm{H}\to\gamma\gamma$ and $\rm \mathrm{H}\to ZZ^{*}\to4\ell$ decay channels. Projections for the High-Luminosity LHC indicate that this uncertainty will be reduced to the level of 20\MeV~\cite{ATL-PHYS-PUB-2018-054, CMS-PAS-FTR-21-007}, mainly driven by the $\mathrm{H}\to\mathrm{ZZ^{*}}\to4\ell$ channel. The measurement is expected to remain limited by statistical uncertainties.

In this Letter, we demonstrate that FCC-ee can measure the Higgs boson mass with a precision of 4\MeV using the recoil-mass technique, which has also been studied at the ILC~\cite{Yan:2016xyx} and CEPC~\cite{An:2018dwb}. This approach exploits the Higgsstrahlung process, \eeZH, which provides a clean and model-independent handle on Higgs boson production. Earlier studies have explored preliminary projections and alternative approaches based on threshold scans at FCC-ee~\cite{Azzurri:2021nmy}. Here, we present the first recoil-mass study combining the electron and muon channels with angular categorization, a comprehensive treatment of the dominant systematic uncertainties, and explicit variations of the detector and accelerator performance.

\begin{figure}[t!]
\centering
\includegraphics[width=0.45\textwidth]{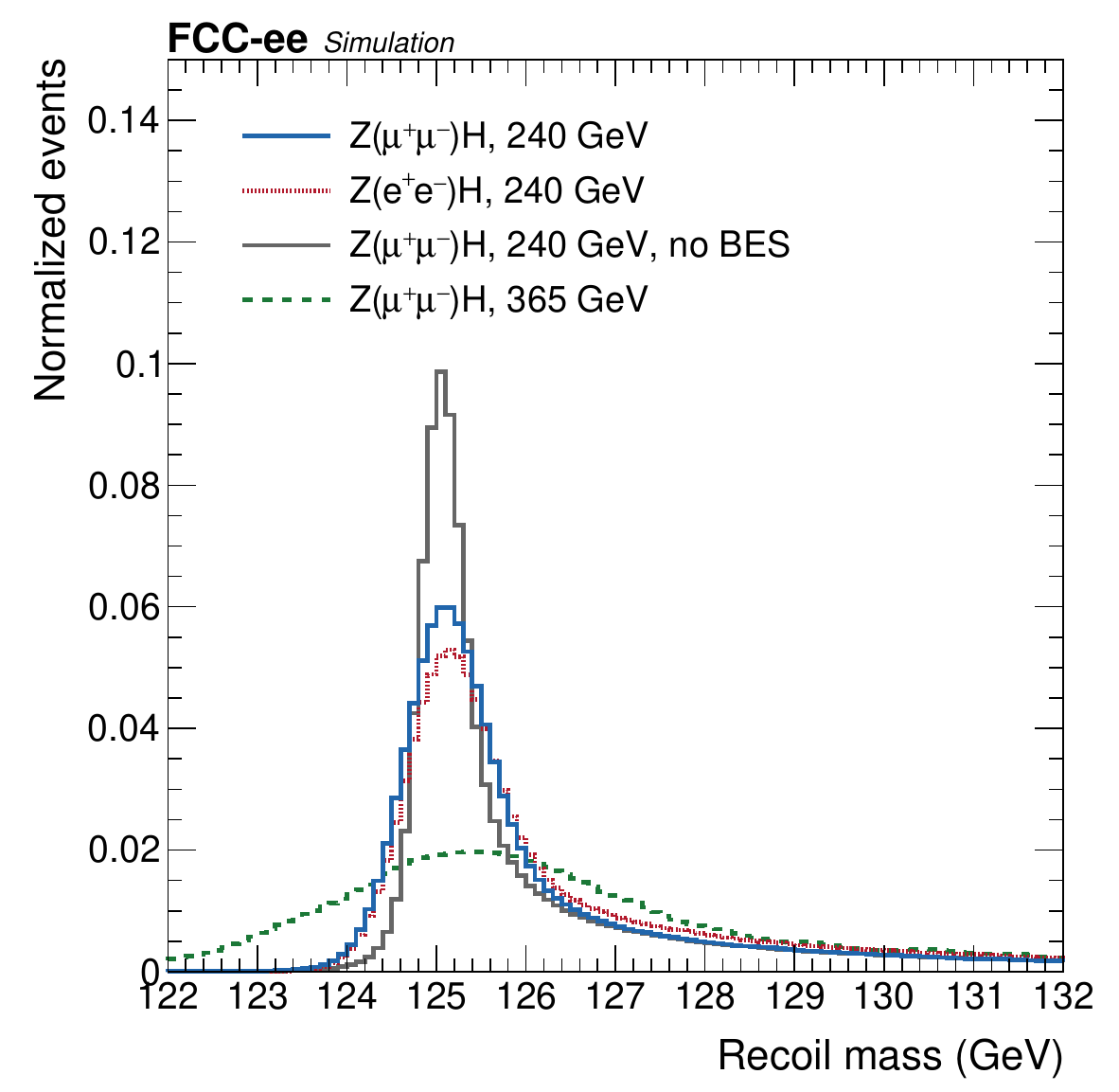}
\caption{Normalized recoil-mass distributions for the muon and electron
channels at \sqrtsZH, compared with the muon channel for zero
beam-energy spread (BES) at \sqrtsZH and for nominal beam conditions at
\sqrtsTop.}
\label{fig:recoilmass}
\end{figure}

At FCC-ee, the dominant Higgs production mechanism is the Higgsstrahlung process, and the baseline physics program includes data taking at a center-of-mass energy of \sqrtsZH, near the maximum of its production cross section, and at \sqrtsTop, around the top-quark pair-production threshold. This topology enables the recoil-mass technique, in which events are selected through the reconstruction of the associated Z boson. Using the precisely known $\sqrt{s}$ and the reconstructed $\rm \mathrm{Z}\to f\bar{f}$ decay, the Higgs boson kinematics is inferred from energy conservation. The recoil mass,
\begin{equation}\label{eq:recoil}
\rm \mrec^2 = (\sqrt{s} - E_{f\bar{f}})^2 - p^2_{f\bar{f}}
= s - 2E_{f\bar{f}}\sqrt{s} + m_{f\bar{f}}^2\,,
\end{equation}
is centered near the Higgs boson mass, where $\rm E_{f\bar{f}}$, $\rm p_{f\bar{f}}$, and $\rm m_{f\bar{f}}$ denote the energy, momentum, and invariant mass of the reconstructed $\rm Z\rightarrow f\bar{f}$ system, respectively. In this analysis, we consider only leptons (electrons and muons) because of the excellent momentum resolution achievable at the proposed FCC-ee detectors. Hadronic and $\tau$-lepton final states are not considered because
of their comparatively poorer experimental resolution. Representative recoil-mass spectra at \sqrtsZHTop are shown in Fig.~\ref{fig:recoilmass}. The asymmetric tail toward high recoil masses originates from initial-state radiation, reducing the collision energy.

Using Eq.~\eqref{eq:recoil} and neglecting the impact of the very precise lepton angle measurements, the recoil-mass resolution is given as
\begin{equation}\label{eq:recoilres}
    \rm \sigma_{\mrec}^2 = \left( \frac{\sqrts}{\mrec} \right)^{2} \sigma_{E_{f\bar{f}}}^{2}  + \left( \frac{\sqrts - E_{f\bar{f}}}{\mrec} \right)^{2} \sigma_{\sqrts}^2,
\end{equation}

where $\rm \sigma_{E_{f\bar{f}}}$ denotes the resolution of the total lepton energy, and $\rm \sigma_{\sqrts}$ represents the center-of-mass energy spread, dominated by beamstrahlung during collisions.

At \sqrtsZH, the recoil-mass resolution is dominated by the lepton momentum resolution, with a smaller contribution from the angular resolution. As shown in Fig.~\ref{fig:recoilmass}, the $\ZmumuH$ channel provides the best performance, while for electrons the resolution is degraded by bremsstrahlung effects. Nevertheless, efficient bremsstrahlung recovery allows the electron-channel sensitivity to approach that of muons, such that the optimal precision on \mh is obtained from their combination.

The recoil-mass resolution directly depends also on the center-of-mass energy spread, $\rm \sigma_{\sqrt{s}}$. For equal and uncorrelated beam-energy spreads, $\rm \sigma_{E_b}$, the two quantities are related by $\rm \sigma_{\sqrt{s}}=\sqrt{2}\,\sigma_{E_b}$. Based on the accelerator parameters in the FCC Feasibility Study Report (FSR)~\cite{FCC:2025uan}, the relative beam-energy spread at \sqrtsZH is $\rm0.185\%$~\cite{FCC:2025lpp}, corresponding to $\rm \sigma_{E_b}=222~\MeV$ and, consequently, $\rm \sigma_{\sqrt{s}}=314~\MeV$. As illustrated in Fig.~\ref{fig:recoilmass}, removing the beam-energy spread from the simulation significantly improves the recoil-mass resolution, therefore improving the statistical precision of the measurement.

Uncertainties in the determination of the average center-of-mass energy, of the beam-energy spread, and of the lepton momentum scale modify the position and shape of the reconstructed recoil-mass distribution and must therefore be controlled below the statistical precision of the measurement, as discussed below.

We do not include the \sqrtsTop running scenario because the predicted precision on \mh improves by only about 1\% when including it~\cite{eysermans_2025_c5dn3-c0s73}. The run plan foresees only a modest integrated luminosity, and the recoil-mass resolution degrades significantly with increasing center-of-mass energy, scaling approximately with $\sqrt{s}^2$. In addition, the larger beam-energy spread, enhanced initial-state radiation, and increased lepton momenta further broaden the recoil-mass distribution, producing a wider signal shape, as shown in Fig.~\ref{fig:recoilmass} (dashed green curve).

Monte Carlo samples and fast detector simulation are used to model FCC-ee operation at \sqrtsZH with an integrated luminosity of 10.8~\si{\invab} corresponding to 4 interaction points. The nominal beam-energy spread introduced above is included in the simulation. Signal and fermion-pair production are generated with \WHIZARDPs~\cite{Kilian:2007gr,Sjostrand:2006za}, while diboson background processes are simulated with \PYTHIAe~\cite{Sjostrand:2014zea}. The detector response is modeled with \DELPHES~\cite{deFavereau:2013fsa} within the \KEYfHEP framework~\cite{key4hep}, using a modified IDEA detector concept~\cite{IDEA1,IDEA2}. It features low-material tracking, a crystal electromagnetic calorimeter, and electron bremsstrahlung recovery based on full-simulation studies, resulting in an electron momentum resolution approximately 25\% worse than that of muons~\cite{eperf}. Alternative detector concepts under consideration are expected to provide similar reconstruction performance~\cite{FCC:2025lpp}, and the impact of concrete variations in the assumed detector performance parameters is discussed below. Dedicated samples are used to evaluate Higgs boson mass and beam-energy-spread uncertainties. Further details of the event generation are provided in
Ref.~\cite{li2025modelindependentzhproductioncross}.

The Higgs boson mass is extracted from the recoil-mass spectrum in \ZH\ events with Z bosons decaying to electrons or muons. The event selection follows the model-independent \ZH\ cross-section
analysis~\cite{li2025modelindependentzhproductioncross}, with minor modifications introduced for this study. The same selection is applied to the muon and electron channels and is designed to suppress the dominant $\WW$, $\ZZ$, and $\Zg$ backgrounds. Candidate events are required to contain at least two oppositely charged leptons of the same flavor and with momentum $\rm p>20~\GeV$. To suppress backgrounds from semileptonic heavy-flavor decays, at least one lepton must satisfy an isolation requirement of $\rm \iso = \Sigma p_{others} / p_{\ell} <0.25$, where the sum runs over the momenta of all other reconstructed particles within a cone of radius $\rm \Delta R < 0.5$ around the lepton direction. If multiple lepton pairs are present, the pair most compatible with a \Z boson is selected by minimizing

\begin{equation}
\rm \chi^2 = A(\mll-\mZ)^2 + B(\mrec-\mh)^2\,,
\end{equation}

where $\rm \mZ=91.2~\GeV$ and $\rm \mh=125~\GeV$. The coefficients $\rm A=0.6$ and $\rm B=0.4$ account for the different resolutions of the dilepton invariant mass and the recoil mass. Final requirements are imposed on the selected lepton pair: $\rm 86<\mll<96~\GeV$, $\rm 20<\pll<70~\GeV$, and $\rm 120<\mrec<140~\GeV$. To further suppress $\Zg$ backgrounds, the missing-momentum direction is required to satisfy $\rm |\costhetamiss|<0.98$. The event-selection cutflows for both lepton channels are included in the supplemental material.

\begin{figure}[t!]
\centering
\includegraphics[width=0.45\textwidth]{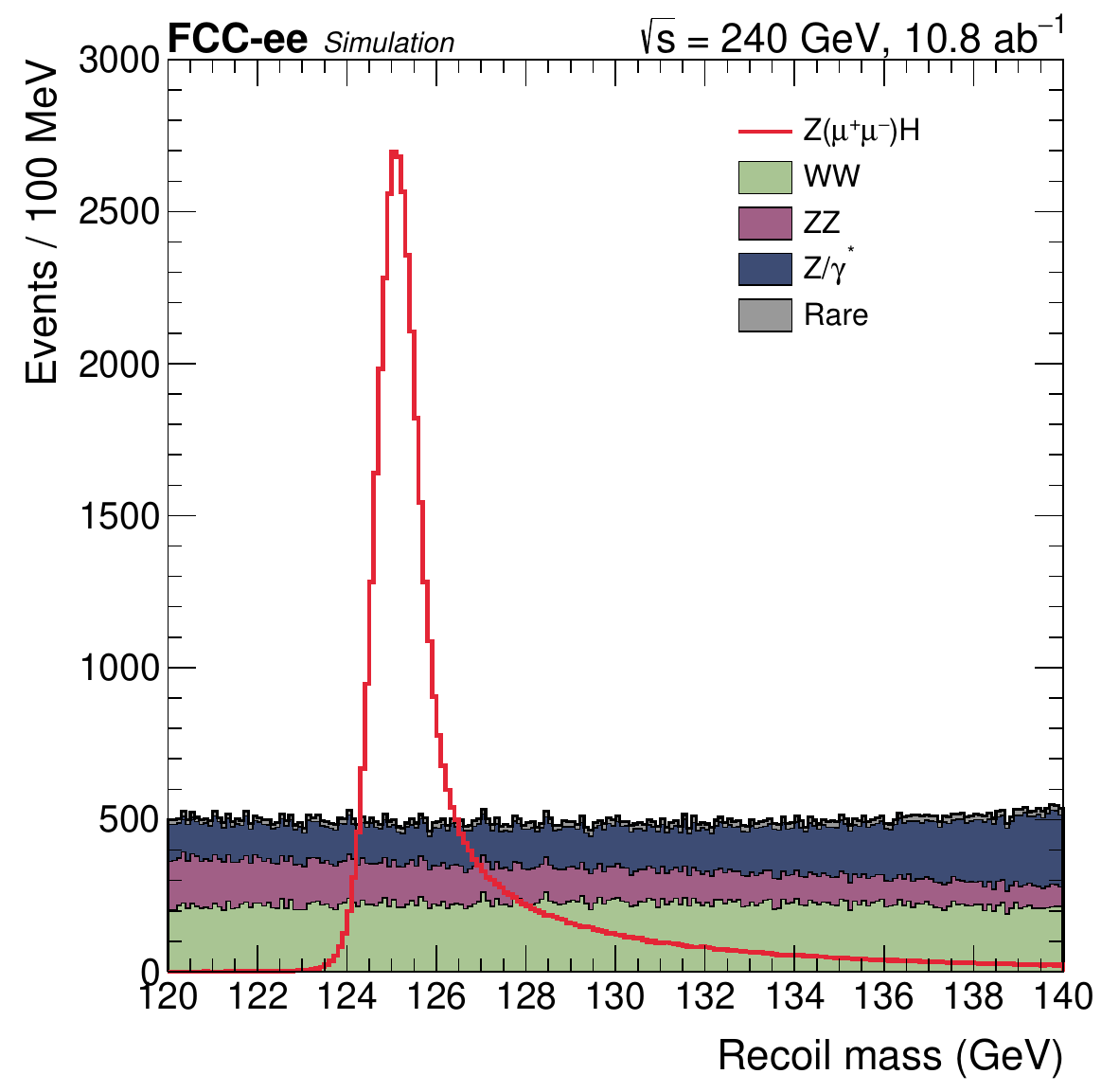}
\caption{Recoil-mass distribution after the full event selection in the muon final state.}
\label{fig:mass:recoildistr}
\end{figure}

After the selection, approximately 66\% of $\ZmumuH$ and 58\% of $\ZeeH$ signal events are retained, corresponding to about $4.8\times10^4$ and $4.5\times10^4$ selected signal events, respectively. The dominant backgrounds are suppressed by 4 orders of magnitude, yielding a final background contribution 5 times smaller than the signal yield in the recoil-mass signal region (Fig.~\ref{fig:mass:recoildistr}). A similar distribution is obtained in the electron channel, although the resolution is degraded by bremsstrahlung and a small contribution from $t$-channel processes, as shown in the supplemental material.

To exploit the variation in recoil-mass resolution across the detector acceptance, events are categorized according to the two lepton polar angles into central--central, central--forward, and forward--forward categories. In total, six categories are used to extract the Higgs boson mass. This categorization improves the sensitivity to \mh by separating regions with different recoil-mass resolutions. The corresponding recoil-mass distributions for the muon categories, illustrating the different resolutions, are provided in the supplemental material.

Recoil-mass distributions are then fitted simultaneously using an unbinned maximum likelihood approach to extract the Higgs boson mass. The signal shape is modeled with Crystal Ball-based functions, while Bernstein polynomials describe the smooth background spectrum. The dependence on \mh is determined from dedicated simulated signal samples. The fit is performed within the \texttt{Combine} framework~\cite{CMS:2024vfx}, allowing \mh and overall signal and background normalizations to float freely. Additional details on the signal parameterization are provided in the supplemental material.

Table~\ref{tab:results} summarizes the statistical precision on \mh obtained in the individual channels and their combination. Combining the muon and electron channels yields a statistical uncertainty of 3.1\MeV. Although the sensitivity is dominated by the muon channel, the combination with the electron channel improves the precision by approximately 22\% despite its reduced recoil-mass resolution and its larger background contribution.

\begin{table}[ht]
\centering
\caption{Summary of the Higgs boson mass precision in the muon, electron, and combined channels, including statistical and systematic uncertainties.}
\label{tab:results}
\begin{tabular}{lll}
\toprule
Channel                 & Stat. only (MeV) & Stat. + syst. (MeV)\\
\midrule
Muon                    & 3.9            & 4.7 \\ 
Electron                & 5.0            & 5.7 \\ 
Combination             & 3.1            & 4.0 \\ 
\bottomrule
\end{tabular}
\end{table}

\begin{figure}[t!]
\centering
\includegraphics[width=0.45\textwidth]{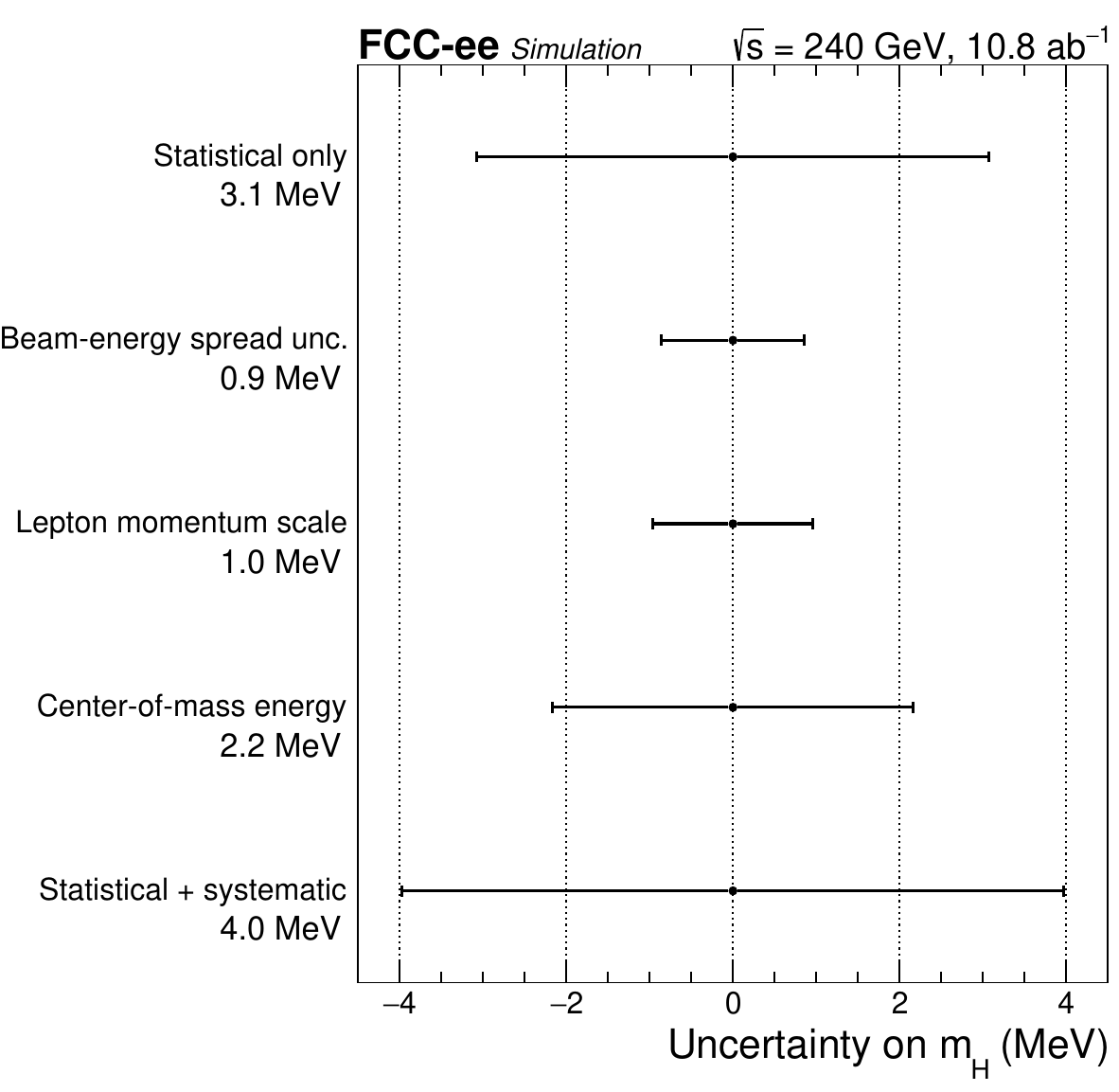}
\caption{Uncertainty on \mh, showing the statistical uncertainty, individual systematic contributions evaluated separately, and their combined effect.}
\label{fig:scan_results}
\end{figure}

The dominant experimental and accelerator-related systematic uncertainties have been evaluated and included in the Higgs boson mass extraction. Their impact is determined using dedicated variations of the recoil-mass templates, propagated through the likelihood fit as nuisance parameters. The largest contribution arises from the determination of the average center-of-mass energy, which enters directly in the recoil-mass calculation, Eq.~(\ref{eq:recoil}). Conservatively assuming an uncertainty of 2\MeV on \sqrts from data-driven measurements of fermion-pair production~\cite{blondel2019polarizationcentreofmassenergycalibration}, the corresponding uncertainty in \mh is estimated to be 2.2\MeV. Additional contributions arise from the uncertainty in the nominal beam-energy spread and the lepton momentum scale, both of which can be constrained in situ using abundant radiative-return dimuon events. A target relative uncertainty of 1\% on the determination of the beam-energy spread, corresponding to an uncertainty of 2.2\MeV on the nominal spread $\rm \sigma_{E_b}$ (222\MeV), is assumed, yielding a 0.9\MeV contribution to the uncertainty on \mh, while the expected $10^{-5}$ constraint on the lepton momentum scale contributes 1.0\MeV~\cite{eysermans_2025_c5dn3-c0s73}. Using a more conservative 6\% relative uncertainty on the determination of the beam-energy spread, obtainable from accelerator instrumentation, increases the former contribution only to 1\MeV~\cite{eysermans_2025_c5dn3-c0s73}. The modeling of initial-state radiation and residual background-shape uncertainties are found to have a negligible impact on the extracted Higgs boson mass. Including all systematic effects increases the overall uncertainty in \mh from 3.1 to 4.0\MeV, dominated by the statistical component. A summary of the results is provided in Table~\ref{tab:results}, and the corresponding uncertainty contributions are shown in Fig.~\ref{fig:scan_results}.

\begin{table}[h]
\centering
\caption{Statistical uncertainty in the Higgs boson mass for various detector and fit configurations, combining the muon and electron channels unless stated otherwise.}
\label{tab:mass:fit_results}
\begin{tabular}{lll}
\toprule
Fit configuration                               & Uncertainty (MeV) \\
\midrule
Default (2~T)                                   & 3.1 \\
Ideal momentum resolution                       & 2.4 \\
Silicon tracker ($\rm \approx 6\%~X_0$)         & 3.9 \\
Magnetic field (3~T)                            & 2.5 \\
Electron resolution (2 $\rm \times$ muon)       & 3.2 \\
Default without beam-energy spread              & 1.7 \\
\bottomrule
\end{tabular}
\end{table}

\begin{figure}[t!]
\centering
\includegraphics[width=0.45\textwidth]{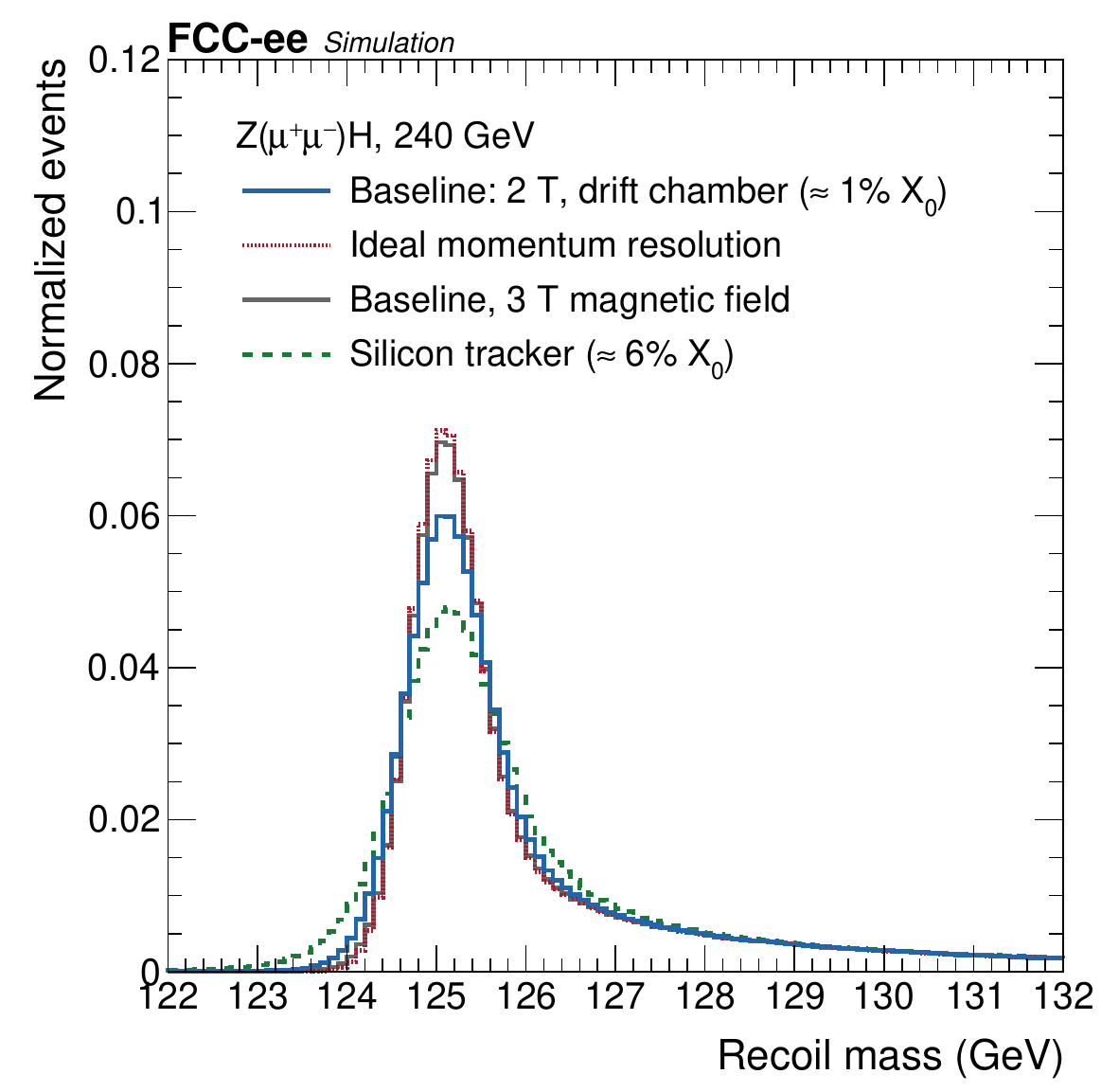}
\caption{Impact of representative variations in tracking-detector performance on the normalized recoil-mass distribution in the muon channel.}
\label{fig:mass:recoil-impact}
\end{figure}

To assess the detector-performance requirements for a few-MeV Higgs boson mass measurement, several detector and analysis variations are studied. Since these variations primarily affect the recoil-mass resolution, only statistical uncertainties are considered in the following. The corresponding results are summarized in Table~\ref{tab:mass:fit_results}, while recoil-mass distributions for selected tracking-related effects are shown in Fig.~\ref{fig:mass:recoil-impact} for the muon channel.

\begin{figure}[t!]
\centering
\includegraphics[width=0.45\textwidth]{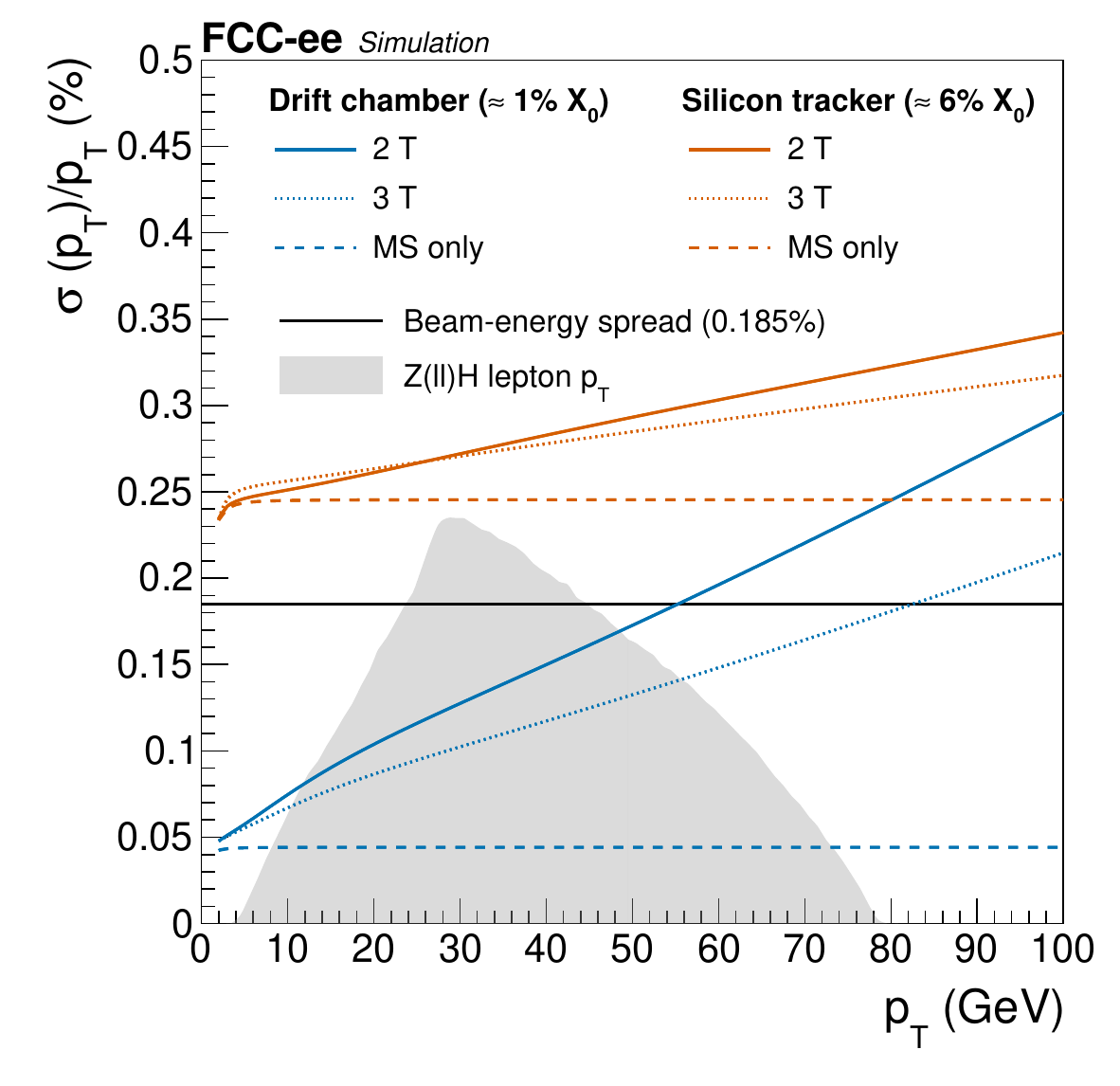}
\caption{Muon transverse-momentum resolution for drift-chamber and silicon-tracker configurations at 2 and 3~T. The total and multiple-scattering (MS) contributions are shown separately. The horizontal line indicates the nominal relative beam-energy spread for comparison, while the shaded distribution shows the lepton $\rm p_{T}$ spectrum in \ZH events at \sqrtsZH.}
\label{fig:tracking_perf}
\end{figure}

As a first benchmark, reconstructed leptons are replaced by their generator-level kinematics after event selection while retaining realistic beam conditions and background contributions. The resulting improvement from 3.1 to 2.4\MeV demonstrates that the nominal detector performance is already close to the intrinsic limitation imposed by the beam-energy spread. The role of the tracking system is investigated by replacing the nominal low-material gaseous tracker with a higher-material-budget conventional silicon tracker~\cite{Bacchetta:2019fmz}. The corresponding momentum resolutions are shown in Fig.~\ref{fig:tracking_perf} as a function of the muon transverse momentum. The figure illustrates the interplay between intrinsic tracking performance and the limitation imposed by the beam-energy spread: while the gaseous tracker, with a material budget of approximately 1\% of radiation length ($\rm X_0$), remains close to the beam-limited regime over the momentum range relevant for \ZH events, the larger material budget of the silicon tracker considered here, about $\rm 6\%~X_0$, leads to a substantially larger multiple-scattering contribution. As a result, the uncertainty in \mh increases by 26\%, from 3.1 to 3.9\MeV. On the other hand, with the gaseous tracking, increasing the magnetic field from 2 to 3~T improves the precision to 2.5\MeV, approaching the ideal-resolution benchmark. Finally, the impact of the electron reconstruction performance is evaluated by degrading the electron momentum resolution from the nominal factor of 1.25 relative to muons to a factor of 2. In this scenario, the combined Higgs boson mass precision degrades only moderately, from 3.1 to 3.2\MeV, indicating that the measurement remains largely driven by the muon channel.

As an accelerator benchmark, setting the beam-energy spread to zero in the simulation improves the statistical precision to 1.7\MeV. Since the beam-energy spread is an intrinsic accelerator quantity, this shows that, once near-optimal tracking performance is achieved, the statistical sensitivity of the recoil-mass technique becomes increasingly limited by the collider beam parameters rather than by the detector resolution.

The results presented here demonstrate the Higgs boson mass precision achievable at FCC-ee using the recoil-mass method in \ZH events with $\rm \Z \to \mu^+\mu^-$ and $\rm \Z \to e^+e^-$ final states at \sqrtsZH, corresponding to an integrated luminosity of 10.8~\si{\invab}. Event selections suppress the dominant backgrounds while preserving signal sensitivity. The recoil-mass distributions are modeled with dedicated analytical parameterizations and fitted simultaneously using a maximum likelihood approach. Categorization according to the lepton polar angle further improves the sensitivity by exploiting variations in the momentum resolution across the detector acceptance.

Including systematic uncertainties, a combined Higgs boson mass precision of 4.0\MeV is achieved. The systematic uncertainty remains 20\% smaller than the expected statistical precision and is dominated by the conservatively estimated center-of-mass energy uncertainty.

The dependence of the measurement precision on detector and accelerator performance has also been investigated. The studies identify low-material tracking, excellent electron reconstruction with efficient bremsstrahlung recovery, and precise beam-energy calibration as key ingredients to fully exploit the statistical potential of the recoil-mass method. The electron channel provides a significant improvement to the combined precision, while studies of detector-performance variations show that the nominal detector configuration already operates close to the beam-limited regime. Once excellent tracking performance is achieved, reducing the beam-energy spread would increase the signal sensitivity and thereby reduce the statistical uncertainty of the measurement.

A model-independent Higgs boson mass measurement at FCC-ee with a precision comparable to the Higgs natural width is therefore expected. Such a measurement provides quantitative performance targets for both the detector and accelerator designs, eliminates \mh\ as a limiting input to global Higgs and electroweak fits, and paves the way for a unique measurement of the electron Yukawa coupling through resonant Higgs production.

\section*{Acknowledgments}

The work of C. Paus and J. Eysermans is supported by the U.S. Department of Energy, Office of Science, Office of High Energy Physics under contract no. DE‐SC0011939.

The work of A. Li is supported by the U.S. Department of Energy, Office of Science, Office of High Energy Physics under contract no. DE-SC0012704.

\bibliographystyle{elsarticle-num} 
\bibliography{FCCee_HiggsMass_PRL}

\clearpage

\section*{Supplemental Material}

\subsection*{Event selection and cutflow}

\begin{figure}[ht!]
\centering
\includegraphics[width=0.48\textwidth]{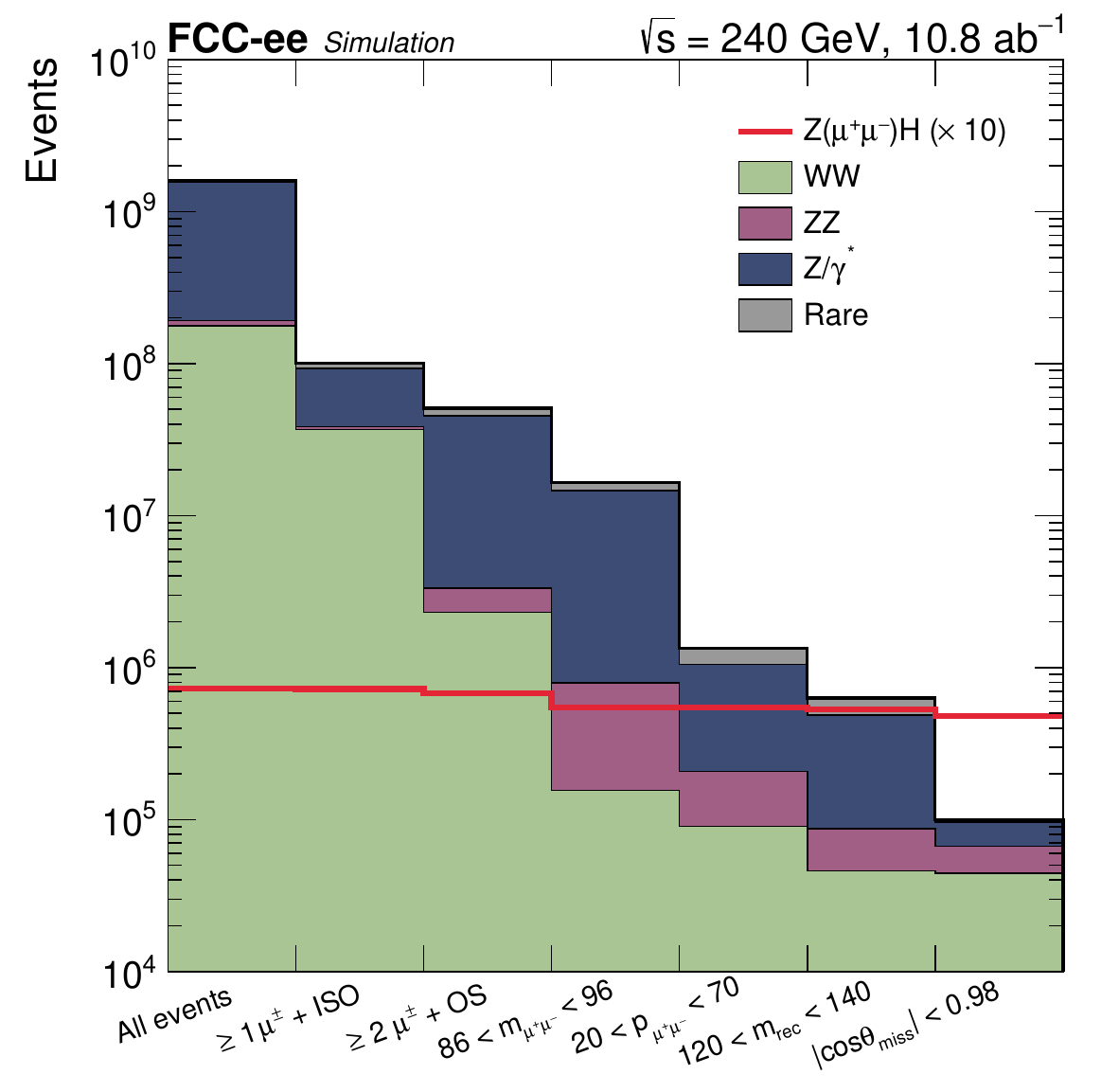}
\includegraphics[width=0.48\textwidth]{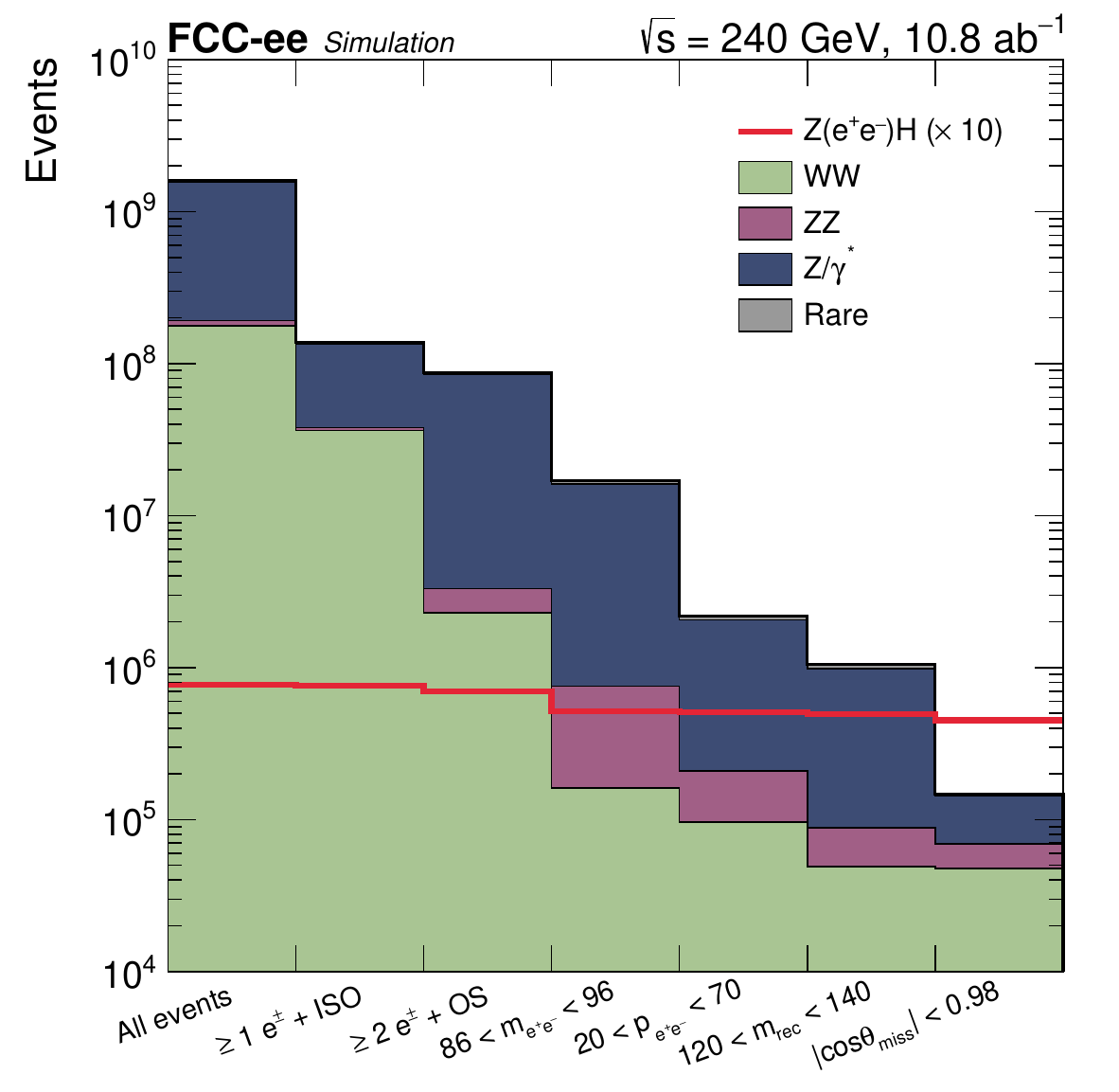}
\caption{Event-selection cutflows for the muon (top) and electron (bottom) channels, showing the signal and dominant background contributions after each successive selection requirement.}
\label{fig:cutflow}
\end{figure}

The event-selection cutflows for the muon and electron channels, including the evolution of the dominant background contributions through the successive selection requirements, are shown in Fig.~\ref{fig:cutflow}.

\subsection*{Electron channel recoil mass}

\begin{figure}[ht!]
\centering
\includegraphics[width=0.48\textwidth]{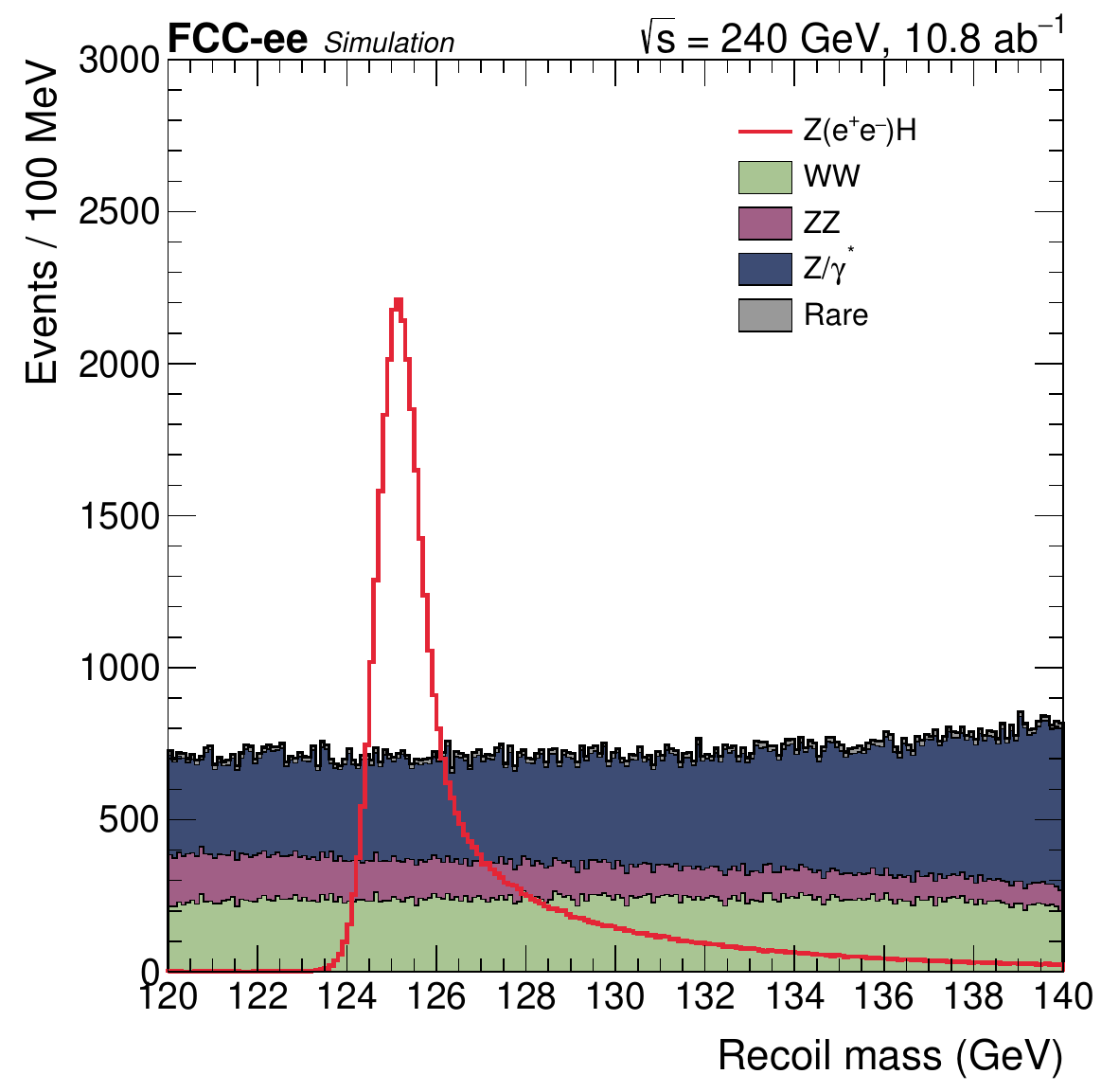}
\caption{Recoil-mass distribution in the electron channel after the full event selection, showing the signal and dominant background contributions.}
\label{fig:recoil_ee}
\end{figure}

The recoil-mass distribution for the electron channel after the full event selection, including the signal and dominant background contributions, is shown in Fig.~\ref{fig:recoil_ee}.

\subsection*{Angular categorization}

\begin{figure}[ht!]
\centering
\includegraphics[width=0.48\textwidth]{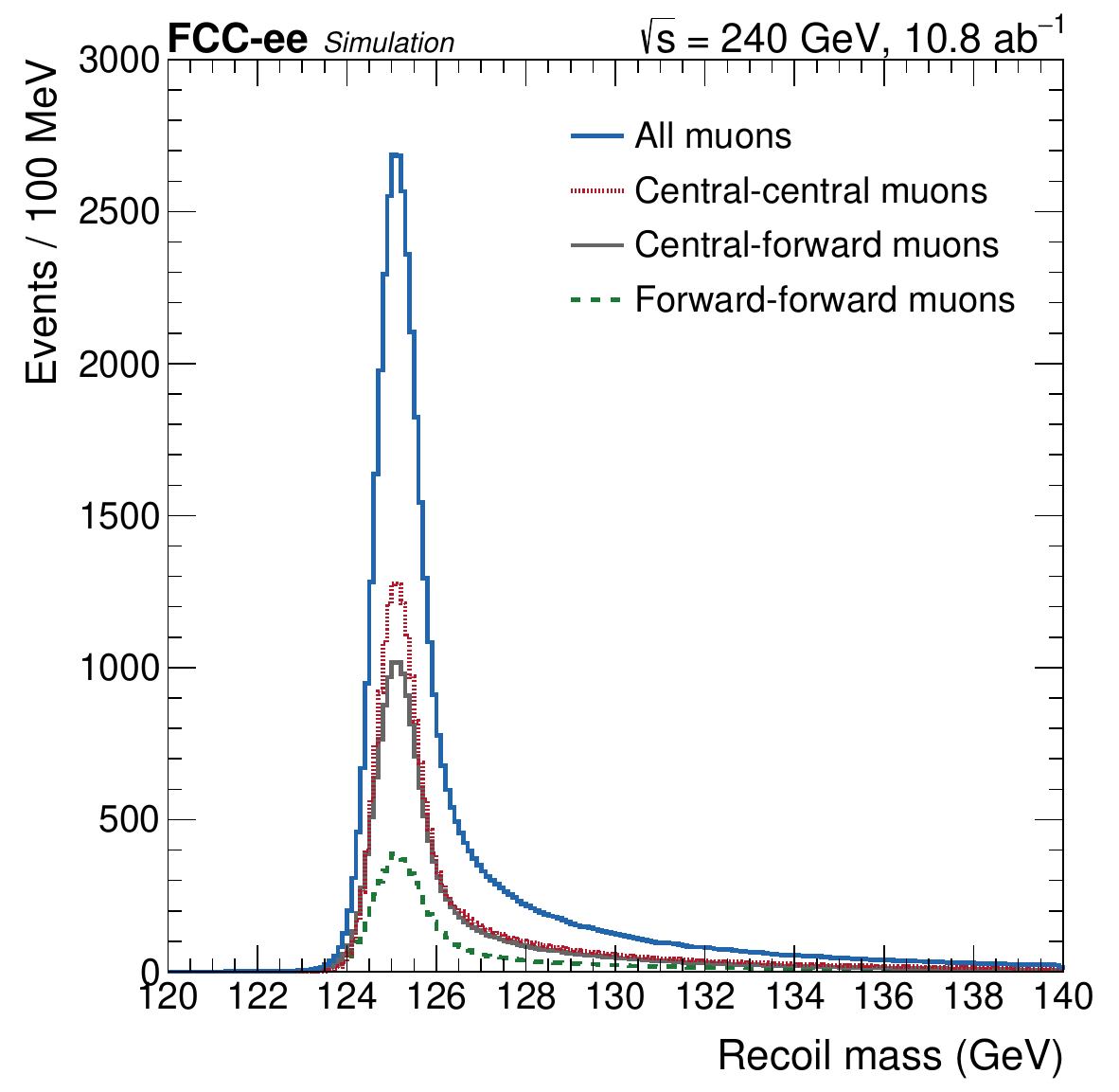}
\caption{Recoil-mass distributions in the muon channel for the central--central, central--forward, and forward--forward angular categories.}
\label{fig:ang_cats}
\end{figure}

The recoil-mass distributions in the different angular categories are shown in Fig.~\ref{fig:ang_cats} for the muon channel, illustrating the variation of the recoil-mass resolution across the detector acceptance. All the categories are fitted simultaneously in the statistical analysis.

\subsection*{Signal and background parameterization}

\begin{figure}[ht!]
\centering
\includegraphics[width=0.48\textwidth]{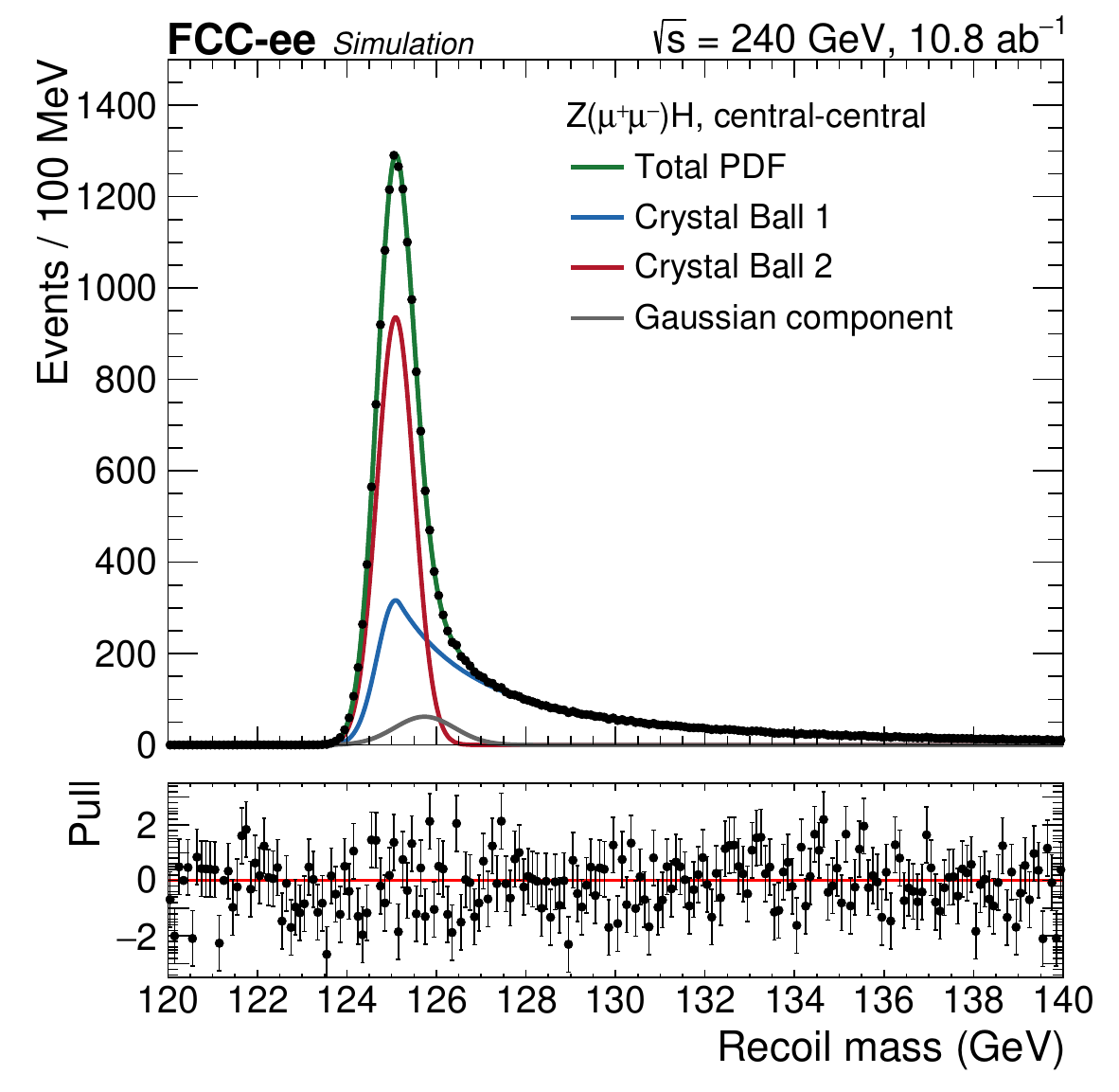}
\caption{Decomposition of the recoil-mass signal model into its two Crystal Ball components and Gaussian component for the muon channel in the central--central angular category.}
\label{fig:sigmodeling}
\end{figure}

The signal recoil-mass distribution is parameterized using an analytic function composed of two Crystal Ball (CB) functions and a Gaussian component:
\begin{align}
\mathrm{pdf}_{\mathrm{rec}} ={}&
c_1\,\mathrm{CB}(\mu,\sigma,\alpha_1,n_1)
+c_2\,\mathrm{CB}(\mu,\sigma,\alpha_2,n_2)
\nonumber\\
&+c_{\mathrm{G}}\,
\mathrm{G}(\mu_{\mathrm{G}},\sigma_{\mathrm{G}}).
\end{align}
This model, containing 11 parameters before normalization constraints, provides an accurate description of the recoil-mass peak and its asymmetric high-mass tail. An example decomposition of the signal model for the muon channel in the central--central angular category is shown in Fig.~\ref{fig:sigmodeling}. The background distributions are modeled using second-order Bernstein polynomials.

\end{document}